\documentclass[aps,prb,twocolumn,groupedaddress,showpacs]{revtex4}

\usepackage{graphicx}

\begin{document}

\title{Chern-Simons Theory for Magnetization Plateaus of Frustrated $J_1$-$J_2$ Heisenberg model}

\author{Ming-Che Chang}
\affiliation{Department of Physics, National Taiwan Normal
University, Taipei, Taiwan}

\author{Min-Fong Yang}
\affiliation{Department of Physics, Tunghai University, Taichung,
Taiwan}
\date{\today}

\begin{abstract}
The magnetization curve of the two-dimensional spin-1/2
$J_1$-$J_2$ Heisenberg model is investigated by using the
Chern-Simons theory under a uniform mean-field approximation. We
find that the magnetization curve is monotonically increasing for
$J_2/J_1 < 0.267949$, where the system under zero external field
is in the antiferromagnetic N\'eel phase. For larger ratios of
$J_2/J_1$, various plateaus will appear in the magnetization
curve. In particular, in the disordered phase, our result supports
the existence of the $M/M_{\rm sat}=1/2$ plateau and predicts a
new plateau at $M/M_{\rm sat}=1/3$. By identifying the onset ratio
$J_2/J_1$ for the appearance of the 1/2-plateau
with the boundary between the N\'eel and the spin-disordered
phases in zero field, we can determine this phase boundary
accurately by this mean-field calculation. Verification of these
interesting results would indicate a strong connection between the
frustrated antiferromagnetic system and the quantum Hall system.
\end{abstract}


\maketitle

Due to quantum and frustration effects, rich physics can
appear in the frustrated quantum spin systems at zero external
field.\cite{frust} Exciting behavior was also observed recently in
several cases with an external magnetic field. For instance, the
recently discovered two-dimensional (2D) $S=1/2$ spin-gap material
SrCu$_2$(BO$_3$)$_2$, which can be described by the
Shastry-Sutherland model,\cite{SS81} exhibits several plateaus at
$M/M_{\rm sat}=1/3$, $1/4$, and $1/8$ in its magnetization curve,
where $M$ ($M_{\rm sat}=1/2$) is the (saturating) magnetization
per site.\cite{Kageyama} The origin of these plateaus and the
nature of the corresponding spin states are under intense
debate.\cite{Miyahara,Momoi,Muller-Hartmann,Fukumoto} Recently, by
mapping onto spinless fermions carrying one quantum of statistical
flux and under a mean-field approximation, Misguich {\it et al.}
show that the original spin model can be related to a generalized
Hofstadter problem, where the spin excitation gaps that produce
the observed magnetization plateaus arise from some of the Landau
level gaps in the integer quantum Hall effect for the fermions on
a lattice.\cite{Misguich} For realistic values of the exchange
constants, they obtain an excellent quantitative fit to the
observed magnetization curve, which demonstrates the success of
their approach.

Another prototype of a realistic frustrated two-dimensional
system, which has been recently realized experimentally in $\rm
Li_2VOSiO_4$ and $\rm Li_2VOGeO_4$ compounds,\cite{carretta} is
the so-called $J_1$-$J_2$ Heisenberg model with the Hamiltonian
\begin{equation}
H =\sum_{\left\langle i,j\right\rangle} J_{ij} \, {\vec
S}_{i}\cdot {\vec S}_{j} -B \sum_i S^z_i ,\label{Hami}
\end{equation}
where the exchange couplings $J_{ij}$ are equal to $J_1$ when $i
,j$ are nearest neighbors on the square lattice; $J_{ij}$ are equal to $J_2$
when $i ,j$ are connected by a diagonal bond. The  external
magnetic field $B$ is applied along the $z$-axis. Both couplings
are antiferromagnetic, i.e. $J_{1,2}>0$, and the spins
$S_{i}=1/2$. This model has been the object of intense
investigation through
years.\cite{Chandra,Read,series1,ed,Schulz,series2,Capriotti,%
lozovik93,Gluzman,Mutter97,ZHP,honecker00,Mutter01} At $B=0$, for
small $J_2/J_1$ where the frustration is weak, the system exhibits
N\'eel ordering described by a wave vector $(\pi ,\pi )$. When
$J_2/J_1$ is large enough, the ground state is dominated by the
next-nearest-neighbor interactions and has a collinear order
described by $(\pi ,0)$ or $(0,\pi )$. There is also a general
consensus on the disappearance of the magnetic ordering at $0.38 <
J_2/J_1 < 0.6$, while the identification of the ground state is
still a subject of much
controversy.\cite{Chandra,Read,series1,ed,Schulz,series2,Capriotti}

Just like the case of the Shastry-Sutherland model, the plateaus
in the magnetization curve are predicted for the $J_1$-$J_2$
model,\cite{lozovik93,Gluzman,Mutter97,ZHP,honecker00,Mutter01}
while the situation is even more controversial. Strong evidence
for a plateau at $M/M_{\rm sat}=1/2$ in the region $0.5 \lesssim
J_2/J_1 \lesssim 0.65$ has been recently reported by Honecker {\it
et al.}\cite{ZHP,honecker00} However, Fledderjohann and
M{\"u}tter\cite{Mutter01} do not find a plateau at $M/M_{\rm
sat}=1/2$ in the region of the quantum disorder phase; but instead
they find some indications for a plateau-like structure at
$M/M_{\rm sat}=2/3$. Most of the previous studies on the plateau
are obtained by the numerical calculations on a small clusters
(the typical  number of lattice sites in these works is about
$6\times6=36$). As discussed in Ref.~\onlinecite{Mutter01}, the
plateau structures in the magnetization curve can depend
sensitively on the system size. Therefore, these results may be
plagued with the finite-size effects. For example, some of the
predicted plateaus may be an artifact of the special lattice
geometry, and the boundary conditions used may frustrate the order
which tends to develop.  Thus the precise determination of the
positions and widths of the plateaus in the frustrated Heisenberg
model is indeed a very delicate problem, and a better theoretical
understanding of the magnetic order and of the mechanisms which
create the plateaus is needed.

To avoid the possible finite-size effects, we apply the
Chern-Simons (CS) theory for the magnetization
plateaus\cite{Misguich} to the 2D spin-1/2
$J_1$-$J_2$ Heisenberg model. Because of its success for the
Shastry-Sutherland model, this approach should give us reasonable
results in the present case. We find that the magnetization curve
is monotonically increasing for $J_2/J_1 < 0.267949$, where the
system at zero field is in the antiferromagnetic N\'eel phase.
Besides, various plateaus will appear in the the magnetization
curve for larger ratios of $J_2/J_1$. In particular, in the
disordered phase, our result supports the existence of the
$M/M_{\rm sat}=1/2$ plateau and predicts a new plateau at
$M/M_{\rm sat}=1/3$. Furthermore, we note that, by
identifying the onset value of $J_2/J_1$ for the appearance of the
1/2-plateau with the critical value of the
phase transition between the N\'eel and the spin-disordered
phases in zero field, this phase boundary, which was
determined earlier by heavy numerical means,\cite{series2,Capriotti}
can be reproduced accurately by this mean-field calculation.

The Hamiltonian in Eq.~(\ref{Hami}) can be rewritten as
\begin{eqnarray}
&&H = H_{xy} + H_z -B \sum_i S^z_i , \nonumber
\\ &&H_{xy} =\frac{1}{2} \sum_{\left\langle i,j\right\rangle} J_{ij}
\, \left( S^+_i S^-_j + S^+_j S^-_i \right), \\ &&H_z =
\sum_{\left\langle i,j\right\rangle} J_{ij} \, S^z_i S^z_j,
\nonumber
\end{eqnarray}
where $H_{xy}$ and $H_z$ are the spin-flip and the Ising parts of
the Hamiltonian at zero field. According to
Ref.~\onlinecite{Misguich}, $H_{xy}$ and $H_z$ are treated in
different ways. First, the Ising part is approximated by a simple
uniform mean-field decoupling,
\begin{equation}
H_z \simeq 4(J_1+ J_2) M \sum_i S^z_i - 2(J_1+ J_2) M^2 N,
\end{equation}
where $N$ is the number of lattice sites. Thus $H_z$ gives a
contribution to the total energy with a simple dependence on the
magnetization. Second, for $H_{xy}$, one can exactly map the spin
operators to spinless fermions attached with a flux tube carrying
one flux quantum of statistical CS magnetic
field,\cite{CS,ywg93,lrf94} where $S^z_i +1/2$ corresponds to the
occupation number $n_i$ of site $i$. Under a mean-field treatment
such that the flux tubes are smeared out into a uniform background
magnetic field, the flux per square plaquette $\phi$ is then tied
to the density of fermions and thus to the magnetization $M$ of
the spin system~:
\begin{equation}
\frac{\phi}{2\pi} = \langle n \rangle = M+{1\over 2} \;.
\label{fluxdensity}
\end{equation}
Because of this flux, each energy band splits to subbands with a
complicated structure. Therefore, the present spin system can be
identified with a Hofstadter problem\cite{hofstadter76} for
fermions moving on a square lattice with nearest-neighbor and
next-nearest-neighbor hoppings.\cite{HK,LCH} For a given $M$ [or
$\phi = 2\pi(M+1/2)$], the mean-field ground state is obtained by
filling the lowest energy subbands with fermions until their
density satisfies $\langle n \rangle=\phi/2\pi$. The one-body
problem from $H_{xy}$ can be straightforwardly analyzed for
rational values of $\phi/2\pi$. For $\phi/2\pi = p/q$ ($p$ and $q$
are mutually prime integers), there are $q$ subbands and the
ground state is the Slater determinant with the lowest $p$
subbands being completely filled. The energy of the filled
subbands leads to another contribution to the total energy. Hence
the total energy per site $E(M)$ as a function of the
magnetization becomes
\begin{equation}
E(M)= {1 \over N} \sum_{\alpha=1,\ldots,p} \sum_{\vec{k}}
\epsilon^{(\alpha)}_{\vec{k}} + 2(J_1+ J_2) M^2.
\label{mean_energy}
\end{equation}
Here $\epsilon^{(\alpha)}_{\vec{k}}$ is the eigenenergy of the
$\alpha$-th subband with the wavevector $\vec{k}$ being restricted
to the magnetic Brillouin zone. The magnetization can be obtained
as a function of $B$ by minimizing $E(M)-BM$. It is clear from
Eq.~(\ref{mean_energy}) that, without the contribution from
$H_{xy}$, the magnetization $M$ will be linearly proportional to
$B$, and there is no magnetization plateau. Therefore, the
appearance of magnetization plateaus is related to certain
features of the Hofstadter spectrum.

\begin{figure}
\centerline{
\includegraphics[width=3.in]{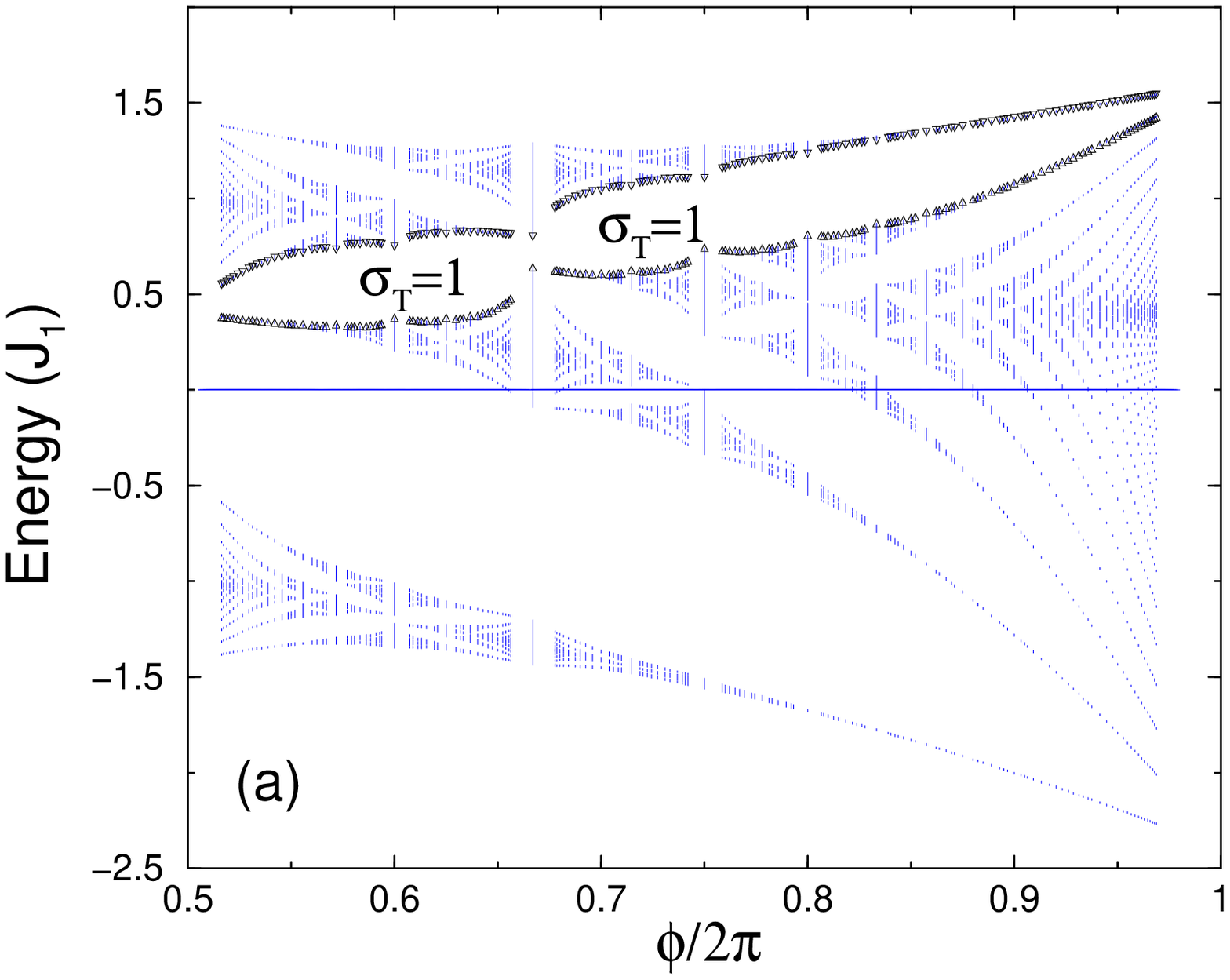}
} \centerline{
\includegraphics[width=3.in]{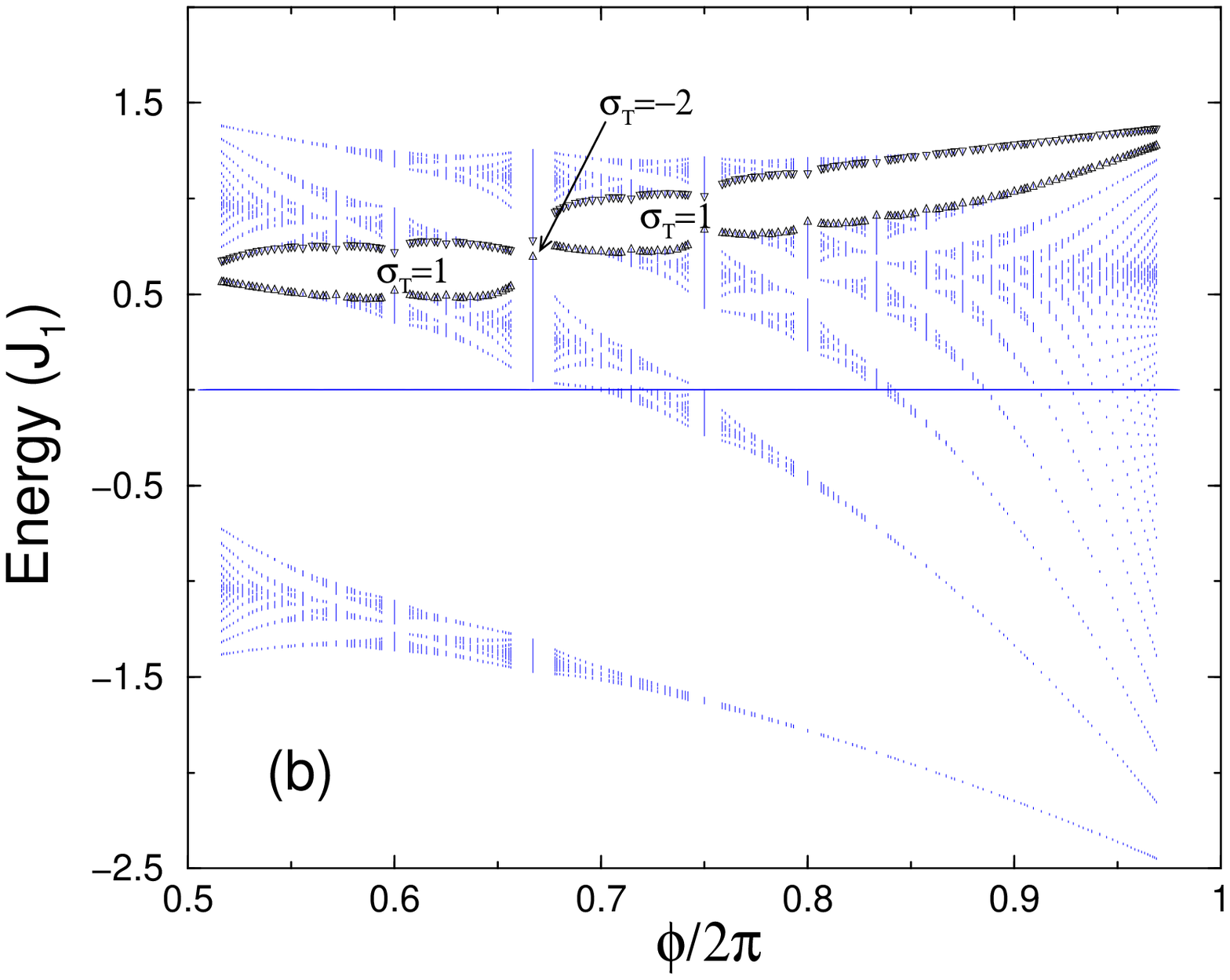}
} \caption{Hofstadter spectra for a square lattice with
nearest-neighbor and next-nearest-neighbor hoppings for $J_2/J_1
=0.2$ (a) and 0.3 (b). Vertical lines mark the energy bands as a
function of the statistical flux $\phi/2\pi$ per square plaquette.
Total Hall conductances $\sigma_{T}$ above the Fermi level from
$M/M_{\rm sat}=0$ ($\phi/2\pi =1/2$) to $M/M_{\rm sat}=1$
($\phi/2\pi =1$) are indicated. The Hall conductance at $\phi/2\pi
= 2/3$ changes from 1 to $-2$ when the upper two subbands touch at
$J_2/J_1=0.267949$.\cite{HK,LCH} } \label{Hof}
\end{figure}

\begin{figure}
\centerline{
\includegraphics[width=3.in]{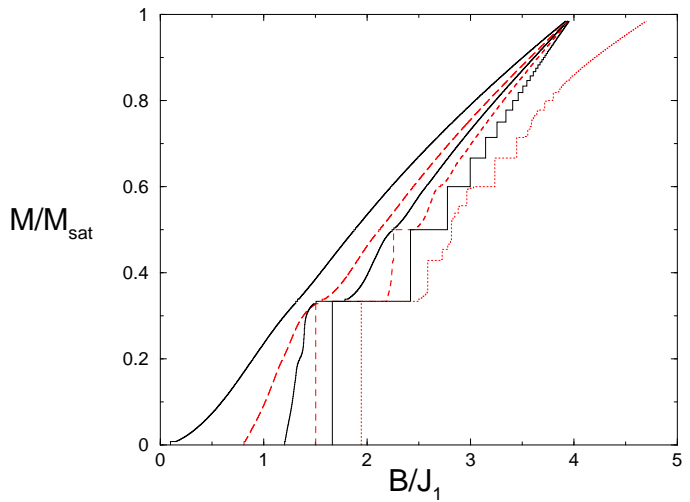}
} \caption{Magnetization curves for the $J_1$-$J_2$ Heisenberg
model calculated using uniform CS mean-field. The curves from left
to right are for $J_2/J_1=0, 0.2, 0.3, 0.4, 0.5$, and 0.7,
respectively. } \label{MB}
\end{figure}

The Hofstadter diagrams for $J_2/J_1$=0.2 and 0.3 are shown in
Fig.~\ref{Hof}, where the lower bold line marks the Fermi level
(highest occupied state) and the upper one marks the lowest
unoccupied level. A jump of the Fermi energy as a
function of $M$ in Fig.~1(b) leads to discontinuity of the slope
of the function $E(M)$. These jumps for various $M$ give rise to
plateaus in the magnetization curve. They are closely related to
the occurrence of band-crossing when the value of $J_2/J_1$ is
varied. For example, in Fig.~1(a) before the upper two subbands
at $\phi/2\pi=2/3$ touch at $J_2/J_1=0.267949$, the Fermi energy
is continuous and there is no magnetization plateau at $M/M_{\rm
sat}=1/3$ (see Fig.~\ref{MB}). However, in Fig.~1(b), the
``pockets" enclosed by the bold lines are separated at
$\phi/2\pi=2/3$. It can be seen that the Fermi level to the right
of the contact point is below (above) the band gap before (after)
band-crossing. As discussed earlier, the Fermi level marks the
position of the $p$-th subband [see Eq.~(\ref{mean_energy})].
Therefore, apparently a subband associated with some flux slightly
larger than 2/3 is shifted above the energy gap after
band-crossing. This shift of a fine subband due to the crossing of the broader subbands at
$\phi/2\pi=2/3$ was studied earlier in the context of the
Hofstadter spectrum.\cite{LCH} It is closely related to the jump
of the integer-valued Hall conductances of the broader
subbands.\cite{LCH,chang} The emergence of the magnetization
plateaus for the spin system thus has an interesting connection
with the change of the integer-valued Hall conductances induced by
band-crossing.

To justify the present approach, it is important to check whether
the plateau states are robust against fluctuations around the
mean-field solutions. It was showed that the Gaussian fluctuations
of the CS gauge field are massless only when the TKNN
integer\cite{tknn82} describing the quantized Hall coefficient of
the fermions on the frustrated lattice becomes {\it
unity}.\cite{ywg93} In that case, the Gaussian fluctuations induce
instabilities for the mean-field ground states. We have computed
the TKNN integers numerically (for example, see Fig.~\ref{Hof})
and found the Gaussian fluctuations to be massive. Therefore, the
plateaus are not destroyed by quantum fluctuations.

Magnetization curves for various $J_2/J_1$ ratios are shown in
Fig.~\ref{MB}. We note that the saturation field $B_{\rm sat}$ can
be computed exactly by identifying the energy $E_f$ of the fully
polarized state with the (exact) minimum energy $E^{\rm min}_{1s}$
of the states with one spin flipped, $E_f=E^{\rm min}_{1s}$. Thus
$B_{\rm sat}/J_1 = 4$ for $J_2/J_1 < 1/2$; $B_{\rm sat}/J_1 = 2
+4J_2/J_1$ for $J_2/J_1 > 1/2$. Our findings agree with these
exact results near the saturation.

In the N\'eel phase, it is expected that the spins cant gradually
from the antiparallel configuration toward the parallel
configuration until the magnetization saturates at the saturation
field $B_{\rm sat}$. The magnetization curve obtained from the
present approach is consistent with this expectation: it is
featureless all the way to full saturation when $J_2/J_1$ is small
(see the curves for $J_2/J_1= 0$ and 0.2 in Fig.~\ref{MB}). Upon
increasing $J_2/J_1$, plateaus emerge and the magnetization curves
become more complex. In particular, because of the band-crossing
at $\phi/2\pi =2/3$ when $J_2/J_1 = 0.267949$ (see Table I of
Ref.~\onlinecite{LCH}), a plateau at $M/M_{\rm sat}=1/3$ is found
[see Eq.~(\ref{fluxdensity})]. This is an unexpected result,
especially for $0.267949<J_2/J_1\le 0.38$ where the ground state
in the absence of an external field is in the N\'eel phase. In the
previous finite-size
studies,\cite{lozovik93,Gluzman,Mutter97,ZHP,honecker00,Mutter01}
there is no indication for the appearance of this plateau.
However, the system sizes and the boundary conditions they used
forbid the appearance of the $M/M_{\rm sat}=1/3$ plateau,
therefore the possibility of this plateau
cannot be ruled out. Furthermore, when $J_2/J_1 = 0.382683$,
another band-crossing in the Hofstadter spectrum occurs at
$\phi/2\pi=3/4$ (see Table I of Ref.~\onlinecite{LCH}), and a
plateau at $M/M_{\rm sat}=1/2$ ensued. It is interesting to find
that the critical value for the appearance of the $M/M_{\rm
sat}=1/2$ plateau agrees quite well with the critical point of the
quantum phase transition at zero field between the N\'eel and the
quantum disorder phases.\cite{series2,Capriotti} This reinforces
our confidence on the present CS mean-field approach. As mentioned
before, while the appearance of a plateau in the quantum disorder
phase had been predicted, the value of the plateau is still under
debate.\cite{ZHP,honecker00,Mutter01} The controversy may come
from the subtle finite-size effects in their investigations. Since
the present CS theory is free from the finite-size effects, we
give a strong support for the existence of the $M/M_{\rm
sat}=1/2$ plateau.

More complex structures in the magnetization curves appear when
$J_2/J_1$ is further increased. For example, when $J_2/J_1 = 1/2$,
a series of plateaus at $M/M_{\rm sat}=n/(n+2)$ is found, which
corresponds to the band-crossing at some magic numbers $\phi/2\pi
=(n+1)/(n+2)$. Irregular plateau structures are found for even
higher values of $J_2/J_1$ (see the $J_2/J_1= 0.7$ curve in
Fig.~\ref{MB}). This behavior is quite similar to the case of the
triangular lattice, where many plateaus are predicted under the
{\em uniform} CS mean-field approximation.\cite{Misguich} In the
case of the triangular lattice, it is shown that, when the
nonuniform mean-field solutions are used, only the main plateau
($M/M_{\rm sat}=1/3$ in that case) survives and other
mini-plateaus disappear. We believe that the same situation will
happen in the present study of the $J_1$-$J_2$ model. That is,
when going beyond the uniform mean-field approximation, many of
the mini-plateaus in the magnetization process may disappear and
only the main plateaus with simple fractions of $M/M_{\rm sat}$
survive. Moreover, as shown in Fig.~\ref{MB}, the spin gap (the
plateau at $M/M_{\rm sat}=0$) opens at $J_2/J_1=0$,
instead of opening at the correct  critical coupling
$J_2/J_1 \sim 0.38$. This feature comes from the fact that the
N\'eel state of the square-lattice antiferromagnet is not
correctly described by the present {\it uniform} mean-field
approximation.

In conclusion, the magnetization curve of the 2D
spin-1/2 $J_1$-$J_2$ Heisenberg model is studied by using the
Chern-Simons theory under a uniform mean-field approximation. In
the disordered phase, our result supports the existence of the
$M/M_{\rm sat}=1/2$ plateau and predicts a new plateau at
$M/M_{\rm sat}=1/3$. Moreover, various plateaus appear in the
magnetization curve both in the disordered and the collinear
phases, which could be an artifact of our mean field
approximation. More work is needed to be conclusive at this point.
We note that it is experimentally accessible to confirm our
results. As mentioned before, the 2D spin-1/2
$J_1$-$J_2$ Heisenberg model with $J_2/J_1 \simeq 1$ has been
recently realized experimentally in $\rm Li_2VOSiO_4$ ($J_1+J_2
\simeq 8.2$ K) and $\rm Li_2VOGeO_4$ ($J_1+J_2 \simeq 8.2$ K)
compounds.\cite{carretta} If the $g$-factor is taken to be 2, the
corresponding saturation fields will be approximately 18 T for $\rm
Li_2VOSiO_4$ and 13 T for $\rm Li_2VOGeO_4$. In both cases, these
values of the magnetic fields can be reached experimentally. Thus
the full magnetization curve could be mapped out to test our
results. Verification of the magnetization plateaus would indicate
a strong connection between the frustrated antiferromagnetic
system and the quantum Hall system.

M.C.C. is supported by the National Science Council of Taiwan
under Contract No. NSC 90-2112-M-003-022. M.F.Y. acknowledges
financial support by the National Science Council of Taiwan under
Contract No. NSC 90-2112-M-029-004.

\end{document}